# Stability of the many-body scars in fermionic spin-1/2 models

Patrice Kolb[1], Kiryl Pakrouski[1]

[1]*Institute for Theoretical Physics, ETH Zurich, 8093 Zurich, Switzerland*

We study the stability of the many-body scars in spin-1/2 fermionic systems under the most typical perturbations in relevant materials. We find that some families of scars are completely insensitive to certain perturbations. In some other cases they are stable to the first order in perturbation theory. Our analytical results apply to a large class of Hamiltonians that are known [1] to support exact many-body scars. For the numerical calculations we choose the deformed $t-J-U$ model that includes both Heisenberg and Hubbard interactions. We propose two new stability measures that are based on physical observables rather than the fidelity to the exact initial wavefunction. They enable the experimental detection of scars and are more reliable from the theoretical and numerical perspectives. One of these measures may potentially find applications in other systems where the exact many-body scars are equally spaced in energy. In small systems and at small perturbations, a regime particularly relevant for quantum simulators, we identify and describe an additional stability exhibited by the many-body scars. For larger perturbation strengths we observe a distinct mode of ergodicity breaking that is consistent with many-body localization.

## I. INTRODUCTION

When an interacting quantum system is initialized to a state that is not an eigenstate of its Hamiltonian this initial state typically quickly disappears (modulo Poincare recurrences that happen at extremely long time scales) in the sense that information stored in it cannot be recovered using local measurements. This expectation is based on the ergodicity assumption that is closely related to the eigenstate thermalisation hypothesis (ETH), the cornerstone conjecture that allows us to connect the microscopic behaviour of quantum systems to the macroscopic statistical physics.

Many-body scars (MBS) are the states for which the above intuition does not hold. Once initialised to a superposition of them the system returns to the exact initial state repeatedly and indefinitely in time preserving the initially stored quantum information. Besides the fundamentally new physics of ETH violation and the potential quantum information applications the MBS in fermionic system have also been shown to possess certain long-range order correlations that may be relevant for superconductivity. While many-body scars have only recently emerged as an interesting research direction the term "scars" in the quantum physics context was first used in 1984 by E. J. Heller [2] to refer to the extra density concentrating around the periodic orbits of one-body chaotic systems and a general method for constructing such eigenstates was proposed in 1994 [3].

Approximate MBS have first been observed experimentally in Rydberg atoms quantum simulators [4] that could effectively be understood [5] as a spin-1/2 chain with a so-called "PXP" Hamiltonian. More recently the equivalent of the PXP model and the scars arising in it have been also implemented in Bose-Hubbard quantum simulators [6] while MBS hosted by the Su-Schrieffer-Heeger model have been identified experimentally on a superconducting qubits quantum simulator [7]. Theoretically MBS have been identified in a great variety of systems [1, 5, 8–48] with Refs. [49–52] providing reviews of the growing body of literature.

In our discussion here we will rely on the group-invariant framework [1, 29, 53] (one of a few [30, 31, 54–57]) that provides a way of describing the general mechanism that leads to the existence of MBS in various systems.

While the conditions a Hamiltonian must satisfy to support perfect MBS are now known, the stability of the effect under generic perturbations is poorly understood. A few such studies exist for the PXP models [58–61], but there are none for the fermionic systems although a large class of fermionic Hamiltonians is known to support ideal MBS [1, 21, 27, 29, 53, 62]. Making progress in this direction is crucial for enabling the experimental studies of MBS in fermionic systems and developing its quantum information processing applications. The group-invariant MBS present in fermionic systems are similar [1, 29, 53] to the "decoherence-free subspaces" [63, 64] which provides a connection to quantum computing literature and makes these systems particularly interesting.

In this work we study the stability of the two families of MBS that naturally occur in a large class of spin-1/2 fermionic models (such as deformed Hubbard or $tJU$ models) under three types of perturbations that may typically be present in materials approximately described by these models: random on-site chemical potential, random on-site magnetic field and randomized density-density interaction.

One of the key properties of the many-body scars is that starting anywhere within the scar subspace the time-evolved wavefunction returns to the exact initial state, the phenomenon known as revivals. This is a consequence of the existence of a common divisor between all the energy gaps separating the scars [29]. For our stability study we focus on the revivals and quantify to which extent the revivals in a perturbed system differ from the exact ones.

We propose several measures of MBS stability beyond the initial wavefunction fidelity. One of these measures is based on the total quadratic Casimir operator of the associated $SU(2)$ symmetry group of the given scar family. In our case it has a natural physical interpretation of the total spin or pseudospin, applies however to any MBS equally-spaced in energy (it has been suggested [57, 65] that actually all the MBS occurring in local models are of this type, see also Ref. [27]). Another measure we propose is based on the particular correlations that are characteristic of the two families of scars and that, crucially, can be measured in experiment.

We obtain a number of analytical results characterising the influence of the considered perturbations on the two MBS families in terms of the measures mentioned above. For some combinations of scar family and perturbation the MBS remain perfectly stable. For some others they are only effected in the 2nd order in perturbation theory. We verify these predictions exactly numerically in finite systems.

At small perturbations and in small systems relevant for potential experiments with quantum simulators we observe and analytically explain additional stability aquired by the MBS.

Under large perturbations for a number of conventionally used measures we observe results that are consistent with the many-body localized [66, 67] phase.

## II. UNPERTURBED SYSTEM

We begin by introducing the general mechanism of many-body scar formation [29] that we will build upon in this work and will refer to as the "$H_0 + OT$ form". Consider a Hilbert space and its subspace $\mathbb{S}$ made of states invariant under a group $G$. The existence of such a subspace is a property of a particular Hilbert space and *not* of the Hamiltonian. By definition any state $|s\rangle$ in this subspace is annihilated by any generator $T_i$ of the group $G$. The same is true also for a sum $\sum_i O_i T_i$, where $O_i$ is any operator such that the product $O_i T_i$ is Hermitian (although scars in non-Hermitian systems exist as well [1]). The requirements (see Ref. [29] for a more strict formulation) on the $H_0$ term is that it does not mix the invariant subspace $\mathbb{S}$ with the rest of the Hilbert space and that all the gaps in its spectrum in the $\mathbb{S}$ subspace have a common divisor. $H_0$ is typically very simple and in this work it is effectively a magnetic field that keeps all the states in $\mathbb{S}$ equally spaced in energy. For any Hamiltonian $H = H_0 + \sum_i O_i T_i$ built to the above prescription the states from $\mathbb{S}$ become scar states because their dynamics is only governed by the simple $H_0$ part while the rest of the Hilbert space has the full Hamiltonian including the terms $O_i T_i$ that can be made strongly interacting and chaotic. The "revivals", the phenomenon of the initial state from $\mathbb{S}$ repeatedly returning to itself is a result of the constructive interference due to the equal energy spacings within $\mathbb{S}$ [29].

The Hilbert space we consider in this work is that of spin-1/2 fermions on a lattice of $N$ sites in any dimension. This case has been studied in detail in Ref. [1] and it was found that up to three families of scar states may appear in a large class of models where the Hamiltonian has the $H_0 + OT$ form. It was also shown that a number of standard interaction terms such as Hubbard can be decomposed as $H_0 + OT$ and thus support scars.

Only two of the three scar families appear if we restrict ourselves to Hamiltonians with real-valued hopping amplitude. In the following we review the properties of these two families originally derived in Ref. [1]. Most of our analytical results are valid for any Hamiltonian (many of them listed in Ref. [1]) that exactly supports the two families as scars. While we expect also the numerical results to be qualitatively model-independent, all our numerical calculations use the same Hamiltonian that was studied numerically in [1] as an unperturbed starting point. This Hamiltonian is specified in the second half of this section.

Each of the two scar families is invariant under its own implementation of a large-rank unitary group which we will denote $U(N)$ and $\widetilde{U}(N)'$, where $N$ is the number of lattice sites. Quadratic hopping terms with real amplitude that describe a free fermionic model are closed under commutation operation. They can be regarded as the generators of these groups and used as the $T_i$ terms of the $H_0 + OT$ form. The exhaustive list of further generators that can be used as $T_i$ for each scar family is given in Ref. [1]. It is the property of this specific Hilbert space [29, 68, 69] that any state that is invariant under one of these large-rank groups automatically has a specific representation with respect to two $SU(2)$ groups. These $SU(2)$ groups are familiar to the condensed matter community as the symmetries of the Hubbard model corresponding to spin (we will call it $SU(2)_{spin}$) and pseudo-spin ($SU(2)_\eta$ in our notation) [70–72].

The scar states include two subspaces with $N + 1$ states each. The first subspace that we will refer to as the "zeta states" is invariant under $U(N)$ and $SU(2)_\eta$ and is spanned by $|n^{\tilde{\zeta}}\rangle$

$$|n^{\tilde{\zeta}}\rangle = \frac{\tilde{\zeta}^n}{2^n \sqrt{\frac{N!n!}{(N-n)!}}} |0^{\tilde{\zeta}}\rangle \, , \quad (1)$$

$$|0^{\tilde{\zeta}}\rangle = \prod_i \frac{c_{i\uparrow}^\dagger + i c_{i\downarrow}^\dagger}{\sqrt{2}} |0\rangle \, ,$$

where $\tilde{\zeta} = Q_3 - iQ_1$ and $|0\rangle$ denotes the vacuum state containing no particles. These states have the highest possible physical spin: they form the spin-$\left(\frac{N}{2}\right)$ representation of $SU(2)_{spin}$ [29] whose generators are

$$Q_A = \sum_{i=1}^N S_i^A \, , \qquad A = 1, 2, 3 \, , \quad (2)$$

with the spin operator at site $i$

$$S_i^A = \frac{1}{2} \sum_{\alpha,\beta} c_{i\alpha}^\dagger \sigma_{\alpha\beta}^A c_{i\beta} = \frac{M_i^A}{2} \, , \quad (3)$$

where $\sigma^A$ are the Pauli matrices, and the Greek indices take two values, $\uparrow$ and $\downarrow$.

The states in eq. (1) are eigenstates of the particular realisation of an $H_0 + OT$ Hamiltonian (7) we are going to consider in this work, however a simpler basis spanning the same subspace exists - see eq. (20). The basis in the scar subspace and the exact expressions for the scar states wavefunctions are fixed by the $H_0$ part of the Hamiltonian one considers. The important physical properties of scars are however qualitatively basis-independent as long as $H_0$ generates a scar spectrum with proportional gaps and preserves the scar subspace [29].

The second family, the "eta states" $|n^\eta\rangle'$, are defined on any bipartite lattice and are invariant under $\widetilde{U}(N)'$ and $SU(2)_{\text{spin}}$; they form the $N+1$ dimensional representation of the pseudo-spin $SU(2)'_\eta$:

$$|n^\eta\rangle' = \frac{(\eta')^n}{\sqrt{\frac{N!n!}{(N-n)!}}}|0\rangle , \quad \eta' = \sum_{j=1}^{N} e^{i\pi j} c_{j\uparrow}^\dagger c_{j\downarrow}^\dagger . \quad (4)$$

These states are also known as the $\eta$-pairing states [73, 74]. In terms of quasiparticles, the $n$th state from this family is an equal superposition of $n$ pairs of spin-1/2 fermions with up and down spin placed on all available sites on the lattice. Therefore each such state has a fixed total particle number: $2n$. In contrast, all the zeta states (1) are located in the half-filling sector with the total particle number $Q = N$.

The eta and zeta states are defined on a lattice of $N$ sites. The only requirement with respect to the spacial arrangement of the lattice sites is that the lattice needs to be bipartite for the eta states. The U($N$) group includes the discrete subgroup of $N!$ permutations $S_N$ of the lattice site indexes. The U($N$)-invariance of the zeta states thus implies invariance under site index transformations. In particular, the group U($N$) includes as its element a transformation where the indexes of all sites are decreased by one and the index of the first site is changed to $N$ (this is also an element of $Z_N \subset S_N \subset U(N)$). In 1D (similar, accordingly modified argument applies in higher dimensions as well) this transformation could be regarded as a spacial translation. Thus we conclude that the zeta states are translation-invariant.

The spacial shift by $a$ sites defined as the index relabelling above is not an element of the $\widetilde{U}(N)'$ (which is a symmetry of the eta states). In contrast, the elements of the $\widetilde{Z}'_N \subset \widetilde{U}(N)'$ are the transformations where the spacial shift is accompanied by the alternating sign change which means the above argument used for zeta states doesn't apply as is. A spacial shift by $a$ sites alone leads to a global sign $(-1)^a$ appearing in front of the operator $\eta'$ (4). Therefore, only the eta states $|n^\eta\rangle'$ (which are eigenstates of the total momentum operator with eigenvalue $\pi n$ [74]) with even $n$ are fully translation-invariant. The states with odd $n$ acquire a global $(-1)^a$ sign under a translation by $a$ sites. In an expectation value of an on-site operator however, the minus signs cancel out which simplifies the way randomised on-site perturbations act on the eta (and zeta) states as discussed in more detail in Sec. IV A.

Another consequence of the U($N$)-invariance is that certain two-point correlators are independent of the distance between the two points when measured in the eta and zeta states [29]. This amounts to the off-diagonal long-range order (ODLRO) present in these states. ODLRO in the eta states was first noticed in Ref. [74] while its derivation as a consequence of the O($N$) $\subset$U($N$)-invariance is given in eq. (18) in the Supplementary Materials of Ref. [29].

The following expectation value of flipping two spins at sites $i$ and $j$ is $(i-j)$-independent for the zeta states

$$G_U(i,j) = \langle c_{i\uparrow}^\dagger c_{i\downarrow} c_{j\downarrow}^\dagger c_{j\uparrow}\rangle . \quad (5)$$

For the eta states the distance-independent expectation value reads [74]

$$G_O(i,j) = \langle c_{i\uparrow}^\dagger c_{i\downarrow}^\dagger c_{j\downarrow} c_{j\uparrow}\rangle . \quad (6)$$

In a number of models a third family of states (eq. 24 in Ref. [1]) closely related to the eta states (4) forms a scar subspace. While we do not consider it in this work explicitly many results obtained for the eta states here can be directly generalised to that related family as well.

As the starting point we use the Hamiltonian that was studied numerically in Ref. [1] where it was shown to support two families of states $|n^{\tilde\zeta}\rangle$ (1) and $|n^\eta\rangle'$ (4) as exact many-body scars. This is a consequence of this Hamiltonian having the $H_0 + OT$ form [29] with respect to the symmetry groups of these two families.

The Hamiltonian is composed of three terms

$$H_h^{tJU} = H^{tJU} + \beta H^b + \gamma Q_2 , \quad (7)$$

where $H^{tJU}$ is the standard $t-J-U$ model, $H^b$ is a symmetry breaking term of the $OT$ form [29] that leaves the scars unchanged and $Q_2$ is a magnetic field used to split the otherwise degenerate $|n^{\tilde\zeta}\rangle$ states: $Q_2|n^{\tilde\zeta}\rangle = (2n-N)|n^{\tilde\zeta}\rangle$) (the $SU(2)_{spin}$-invariant states $|n^\eta\rangle'$ are annihilated by $Q_2$).

In particular, these terms read

$$H^{tJU} = \sum_{\langle ij\rangle\sigma}(tc_{i\sigma}^\dagger c_{j\sigma} + h.c.) + J\sum_{\langle ij\rangle}\vec{S}_i\cdot\vec{S}_j+$$
$$+U\sum_i n_{i\uparrow}n_{i\downarrow} - \mu Q \quad (8)$$

$$H^b = \sum_{\langle ij\rangle} r_{ij}(\tilde{M}_i + \tilde{M}_j)T_{ij} , \quad (9)$$

where $r_{ij} \in [0,1]$ are real random numbers and

$$\tilde{M}_i = r_M c_{i\uparrow}^\dagger c_{i\uparrow} - q_M c_{i\downarrow}^\dagger c_{i\downarrow} , \quad (10)$$
$$T_{ij} = \sum_\sigma c_{i\sigma}^\dagger c_{j\sigma} + h.c. ,$$

with real random numbers $r_M = 1.426974$, $q_M = 2.890703$.

The hopping operators $T_{ij}$ are shown in Ref. [1] to be the generators of the symmetry group of the scar states and therefore they as well as the full term $H^b$ annihilate the scars exactly. The randomness built into $H^b$ allows to break most symmetries of the $t-J-U$ model that are irrelevant for our purposes. The particle number conservation symmetry is however preserved.

The energies of the scar states are given by

$$E_\eta^n = (U-2\mu)n, \quad (11)$$

$$E_{\tilde\zeta}^n = \frac{J}{4}(N-1) - \mu N + \gamma(2n-N), \quad (12)$$

where $n$ is the index of a state in its respective family (4) or (1).

For all numerical results we present we use the parameters

$$t = 1, \; J = 0.1, \; U = 1, \; \mu = 0, \; \beta = 1, \text{ and } \gamma = \frac{1}{4}. \quad (13)$$

They are chosen to place the scars in the middle of the spectrum (see for example Fig. 5a).

For the numerical calculations we perform exact diagonalisation of the Hamiltonians obtaining all their eigenvectors and eigenvalues, allowing us to simulate lattices of up to $N = 9$ sites. With this information at hand we can also perform the unitary time evolution exactly. The starting state for the time evolution is typically an equal-weight superposition of the $N + 1$ states from either scar subspace. Calculations involving the zeta states are performed within the half-filling sector, while some of the calculations with the eta states are in the full Hilbert space. The lattice dimension does not play a role for exact scars [1]. This remains true in presence of two of our single-body perturbations (but might matter for the density-density term). In this work we perform the numerical calculations in 1D and use open boundary conditions such that also the odd-N lattices are bipartite.

### III. PERTURBATIONS

While the MBS are exact for the Hamiltonian (7) the purpose of the present work is to study the stability of the MBS in presence of perturbations that would be most natural in a material or system that is well described by the $t - J - U$ model. The perturbations we consider are the on-site potential

$$\Delta^\mu = \sum_{j=1}^N \lambda_j^\mu n_j, \quad (14)$$

the on-site magnetic field

$$\Delta^B = -\sum_{j=1}^N \sum_{A=X/Y/Z} \lambda_j^{B,A} M_j^A \quad (15)$$

with the magnetization component $M_j^A$ as given in (3), and the density-density interaction

$$\Delta^{dd} = \sum_{j,k} \lambda_{j,k}^{dd} n_j n_k, \quad (16)$$

where $\lambda_j^\mu$, $\lambda_j^{B,A}$ and $\lambda_{j,k}^{dd}$ are site-dependent perturbation strengths, $n_j = \sum_\sigma c_{j\sigma}^\dagger c_{j\sigma}$ - occupation on site $j$.

Note that the analytical perturbative results we will obtain are valid for any (incl. long-range) density-density interactions of arbitrary shape. For our numerical studies we restrict ourselves to the nearest-neighbour interaction that is the most likely perturbation we can expect.

In the numerical calculations the perturbation strengths on each site are randomly drawn and are Gaussian distributed with standard deviation $\lambda$.

The group-invariant states (4) and (1) are exactly annihilated by the generators of their respective symmetry group listed in Ref. [1]. In particular, we will make use of $K_j = n_j - 1$, the generator [1] of the U(N) subgroup of the full symmetry group of the zeta states (1) and $M_j^A$ (3) - the generators of the spin SU(2), one of the symmetries of the eta states.

The action of the on-site potential perturbation on the zeta states is simply

$$\Delta^\mu |n^{\tilde\zeta}\rangle = \sum_{j=1}^N \lambda_j(\mathbb{1} + K_j) |n^{\tilde\zeta}\rangle = \sum_{j=1}^N \lambda_j |n^{\tilde\zeta}\rangle. \quad (17)$$

The magnetic field perturbation term annihilates the eta states: $\Delta^B |n^\eta\rangle' = 0 |n^\eta\rangle'$.

And the density-density interaction has a simple action on the zeta states:

$$\Delta^{dd} |n^{\tilde\zeta}\rangle = \sum_{j,k} \lambda_{j,k}^{dd}(K_j + \mathbb{1})(K_k + \mathbb{1}) |n^{\tilde\zeta}\rangle$$
$$= \sum_{j,k} \lambda_{j,k}^{dd} |n^{\tilde\zeta}\rangle \quad (18)$$

In all the three cases above the perturbations shift all the energies within a scar family (eta or zeta) by a constant which leaves their gaps unchanged and therefore has no effect on the revivals within each family.

### IV. ANALYTICAL ANALYSIS

The revivals of any initial state from the scar subspace are a consequence [1, 29] of constructive interference which follows from the existence of a common divisor of the energy gaps separating the scar states for the Hamiltonian (7) we start with. Here we study analytically how this property is affected by the perturbations.

#### A. Perturbation theory

In order to characterize the effect of perturbations on the revivals in the remaining cases with non-trivial action we use stationary and time-independent perturbation theory and find the lowest order breaking the equal spacings. The first order correction to the energy of a normalized state $|\psi\rangle$ due to an on-site perturbation $\lambda_i V_i$ with $\lambda_i \in \mathbb{R}$ is given by $\Delta E^\psi = \lambda_i \langle\psi|V_i|\psi\rangle$ and we evaluate it below analytically. For both scar families the expectation value $\langle\psi|V_i|\psi\rangle$ does not depend on the site index $i$. For the zeta states it is a consequence of the translation invariance discussed in Sec. II. For the eta states, we first notice that for a suitable $\alpha$, translating any on-site operator $V_i$ gives the same operator acting on the nearby site $V_{i+1}$: $V_{i+1} = e^{i\alpha\hat{p}} V_i e^{-i\alpha\hat{p}}$, where $\hat{p}$ is the momentum operator that is a generator of translations. Now, for any eta state we have $\langle n^{\eta'}|O_{i+1}|n^{\eta'}\rangle = \langle n^{\eta'}|e^{i\alpha\hat{p}} O_i e^{-i\alpha\hat{p}}|n^{\eta'}\rangle = e^{i\alpha\pi n} e^{-i\alpha\pi n} \langle n^{\eta'}|O_i|n^{\eta'}\rangle = \langle n^{\eta'}|O_i|n^{\eta'}\rangle$, where we used the fact that every eta state is an eigenstate of the momentum operator [74].

Because the considered expectation value is site-independent the first order correction due to $\sum_i \lambda_i V_i$ is given for both scar families by $\left(\sum_i \lambda_i\right) \langle \psi | V_i | \psi \rangle$.

### 1. Random potentials and $|n^\eta\rangle'$ scars

Consider the single-site perturbation $\lambda_l^\mu n_l$. The average occupation of each site in an eta state $|n^\eta\rangle'$ is equal (see eq. (4)) to $\frac{2n}{N}$. This already suggests that the first order correction could be

$$\Delta E_n^{\eta'}(\lambda_l^\mu) = \langle n^\eta |' \lambda_l^\mu n_l | n^\eta \rangle' = \frac{2n}{N} \lambda_l^\mu. \quad (19)$$

We show in Sec. A 1 that this expression is indeed exact. Because the correction is linear in the index of the scar state in the tower the $|n^\eta\rangle'$ scars remain equally spaced in the first order with the modified gap $\Delta^{1,\eta} = \Delta^\eta + \frac{2 \sum_l \lambda_l^\mu}{N}$.

### 2. Random fields and $|n^{\tilde{\zeta}}\rangle$ scars

Consider a field of a fixed direction $A = X/Y/Z$ applied to the site $l$: $-\lambda_l^{B,A} M_l^A = -2\lambda_l^{B,A} S_l^A$. We can use the fact [1] that the $|n^{\tilde{\zeta}}\rangle$ states in (1) can be obtained by a rotation around the $X$-axis from the simpler basis $|n^\zeta\rangle$

$$|n^\zeta\rangle = \frac{\zeta^n}{\sqrt{\frac{N!n!}{(N-n)!}}} |0^\zeta\rangle, \qquad |0^\zeta\rangle = \prod_{j=1}^N c_{j\downarrow}^\dagger |0\rangle, \quad (20)$$

where $n = 0, \ldots, N$, and

$$\zeta = Q_1 + iQ_2 = \sum_{j=1}^N c_{j\uparrow}^\dagger c_{j\downarrow} \quad (21)$$

is the spin raising operator.

The expectation values of interest can be expressed in the simpler basis as

$$\langle n^{\tilde{\zeta}} | S_l^X | n^{\tilde{\zeta}} \rangle = \langle (N-n)^\zeta | S_l^X | (N-n)^\zeta \rangle \quad (22)$$
$$\langle n^{\tilde{\zeta}} | S_l^Y | n^{\tilde{\zeta}} \rangle = -\langle (N-n)^\zeta | S_l^Z | (N-n)^\zeta \rangle$$
$$\langle n^{\tilde{\zeta}} | S_l^Z | n^{\tilde{\zeta}} \rangle = \langle (N-n)^\zeta | S_l^Y | (N-n)^\zeta \rangle$$

The states $|n^\zeta\rangle$ (20) only include configurations with $n$ up spins and the remaining $N-n$ down spins. The operators $S_l^X$ and $S_l^Y$ flip spins, change the ratio between up and down spins and therefore lead to vanishing expectation values on the right hand side of (22).

The only non-vanishing first order contribution is due to the second equation in (22). Here we note that the $n$ up spins in the state $|n^\zeta\rangle$ are evenly distributed over the lattice. The on-site spin expectation value in $Z$ direction is thus given by

$$\langle n^\zeta | S_l^Z | n^\zeta \rangle = \frac{1}{2} \frac{n}{N} - \frac{1}{2} \frac{N-n}{N} = \frac{n}{N} - \frac{1}{2}, \quad (23)$$

|  | $|n^{\tilde{\zeta}}\rangle$ | $|n^\eta\rangle'$ |
|---|---|---|
| On-site potential | – | 2nd order |
| On-site field | 2nd order | – |
| Density-density | – | 1st order |

TABLE I: For each considered perturbation the lowest order of the energy perturbation which breaks the equal spacings is shown. A dash (–) indicates that the perturbation does not change the energy spacings at all. For the cases when only the first order perturbation preserves the equal spacings we verified numerically that the next, second order actually perturbs them.

which leads to the expectation value in the $|n^{\tilde{\zeta}}\rangle$ basis

$$\langle n^{\tilde{\zeta}} | S_l^Y | n^{\tilde{\zeta}} \rangle = \frac{n}{N} - \frac{1}{2}. \quad (24)$$

The constant shift does not change the gap while the linear in $n$ term again keeps the $|n^{\tilde{\zeta}}\rangle$ scar states equispaced, changing their gap to $\Delta^1 = \Delta + \frac{\sum_{i=1}^N \lambda_i^{B,Y}}{N}$.

### 3. Nearest-neighbour density-density interaction and $|n^\eta\rangle'$ scars

In Sec. A 2 of the Appendix we show that the first order correction to the scar energies has quadratic dependence on the state index in the tower for density-density interactions

$$\lambda_{l,m}^{dd} \langle n^\eta |' n_l n_m | n^\eta \rangle' = 4\lambda_{l,m}^{dd} \frac{n(n-1)}{N(N-1)} \quad (25)$$

and owing to the high symmetry of the scar states doesn't depend on the site indexes $l, m$.

The equal energy spacing is broken already in the first order and the gap now becomes dependent on $n$: $\Delta^1(n) = U - 2\mu + 4 \left(\sum_{l,m} \lambda_{l,m}^{dd}\right) \frac{2n}{N(N-1)}$. There is however one special line in the parameter space $U - 2\mu = k \cdot \Delta\epsilon$ with $k \in \mathbb{Z}$ and $\Delta\epsilon = \frac{4 \sum_{l,m} \lambda_{l,m}^{dd}}{N(N-1)}$ where the modified energy gap of the $n$th scar state equals $(k + 2n) \cdot \Delta\epsilon$ and all the gaps are therefore integer multiples of $\Delta\epsilon$. The existence of such a common divisor between all the scar gaps leads to revivals with the period $T = \frac{2\pi}{\Delta\epsilon}$ that are preserved in first order for this special choice of $U$ and $\mu$.

Table I summarizes our analytical results for all the combinations of the scar states and perturbations. We stress that in three cases the perturbation has no effect on the scars and their exact revivals continue indefinitely also in the perturbed Hamiltonian.

## B. Quality factor

If the unperturbed system is initialised to an arbitrary state within the scar subspace, the fidelity, the overlap between the initial and time-evolved state, exhibits non-decaying oscillations [1] with period $T$. An example of this behaviour is illustrated in Appendix Fig. 6.

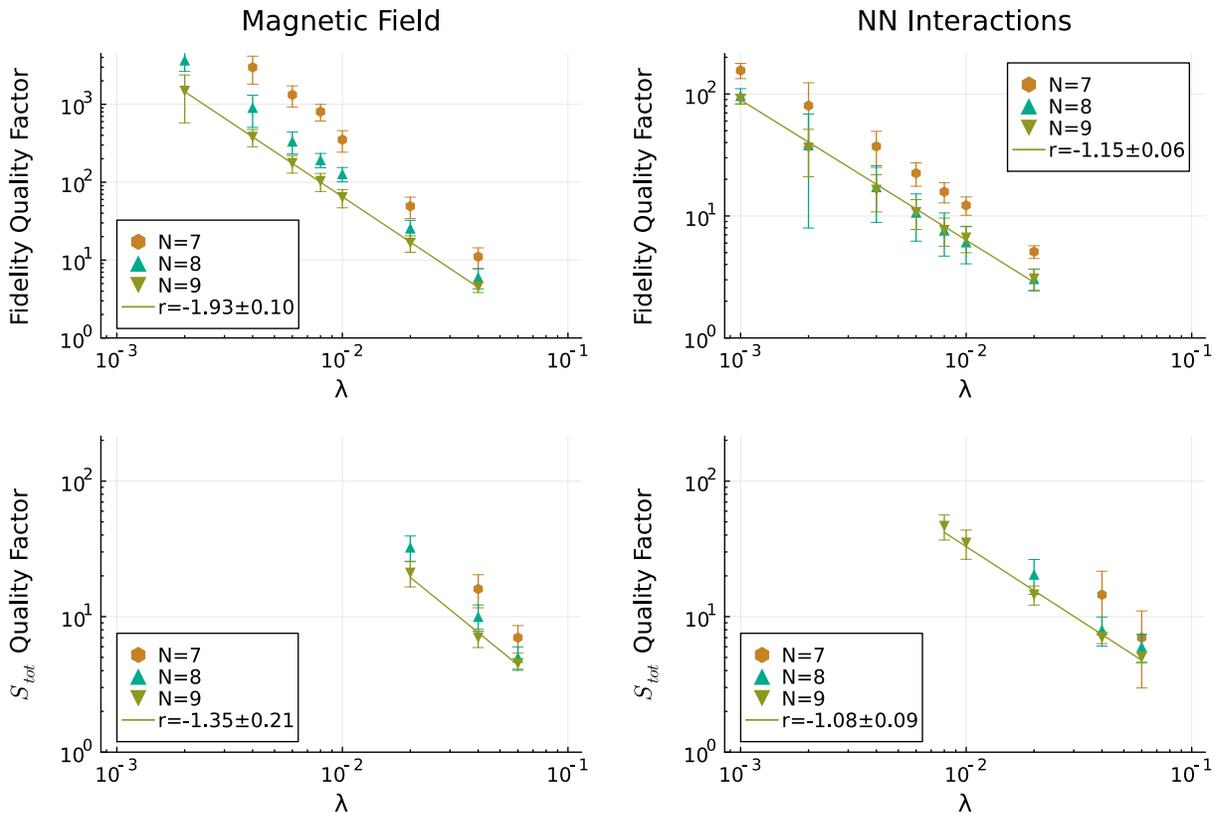

FIG. 1: Dependence of the quality factor on the perturbation strength $\lambda$ for finite system sizes. In the top row (panels a) and b)) the quality factor characterises the initial state fidelity. In the bottom row the quality factor is based on the total spin (panel c)) or pseudo-spin (panel d)). For the figures in the left column (a),c)) the initial state is in the zeta subspace (1) (product, "all spins up" state). The perturbation is the random on-site magnetic field. For the figures in the right column the initial state is in the eta subspace (4). The perturbation is the nearest-neighbour density-density interaction. The fitted exponents $r$ for $N=9$ are indicated. In all cases, for each value of $\lambda$ we average over 10 realizations of the perturbation. The medians and their bootstrapped standard deviations are shown. The threshold is set to 0.75. Datapoints where in more than half of the randomized perturbation realizations the fidelity does not peak below the threshold during the simulated time are omitted. This limits the displayed $\lambda$ range from below.

We use the *quality factor* as one of the measures characterising the decaying oscillations in presence of imperfections. We define the quality factor as the number of oscillations that occur before their amplitude decreases below a certain threshold to be specified. For observables which do not oscillate we instead define the quality factor as the ratio $\frac{t_d}{T}$, where $t_d$ is the time after which the quantity of interest decreases below the threshold.

The dependence of the fidelity quality factor on the perturbation strength can be understood analytically. In Sec. B of the Appendix we use time-dependent perturbation theory to show that at small perturbation strengths $\lambda$ and a large enough threshold the quality factor has a power law dependence on $\lambda$

$$Q \propto \lambda^{-r}, \qquad (26)$$

where $r = 2$ if the first order energy perturbation leaves the scar states equispaced and $r = 1$ otherwise, therefore for a particular scar family-perturbation type pair $r$ can be read off the Table I. Just like the results in Table I the quality factor dependence (26) is thus valid in any model of the $H_0 + OT$ form [1] where the eta (4) or zeta states (1) are scars.

The top row in Fig. 1 shows that the expected power law (26) does approximately describe the actual dependence of the fidelity quality factor for small perturbations $\lambda \leq 0.1$ (magnetic field) and $\lambda \leq 0.01$ (density-density and on-site potential, see Appendix D). Here and throughout the paper we use the "bootstrapped standard deviation" as an estimate of the error bar. We collect the data for various realisations of the random weights that enter the definition of the perturbation part of the Hamiltonian (see Sec. III). In order to determine the bootstrapped standard deviation, we take $B$ random samples containing $P$ elements each from the data. The samples are taken with replacement, i.e. we may pick the same datapoint multiple times. For each sample $i$ we determine the median $m_i$. The bootstrapped standard deviation of the median of the full data is then given by the estimated standard deviation of the sample medians: $\sigma_B = \sqrt{\frac{1}{B-1}\sum_1^B (m_i - \bar{m})}$ with $\bar{m}$ the average of all $m_i$. We used $B = 100000$ and $P$ as half the amount of available data.

## V. FURTHER STABILITY MEASURES

The quality factor based on fidelity is a good theoretical measure of the revivals stability. In experiment however, measuring the overlap with the initial state may not be possible. We define here several other measures that capture similar information as the fidelity-based quality factor but at the same time are based on experimentally measurable quantities.

Both families of scars we consider are the maximum-spin representation of an SU(2) group. For the eta states (4) this is the pseudo-spin SU(2) while for the zeta states (1) it is the regular spin SU(2). In either case the scar family forms a basis in a subspace where the quadratic Casimir (total angular momentum) of the respective SU(2) has the maximum possible value of $N/2$. This value remains unchanged for any state that starts anywhere in the subspace and then evolves with the unperturbed (deformed) $t-J-U$ Hamiltonian (7). In presence of a perturbation the time-evolved state will attain some weight outside of the scar subspace. This can be detected by the decrease of the quadratic Casimir operator measured in this state. The relevant generators of the spin SU(2) are given by $L_+ = \tilde{\zeta}$ (see eq. (1)), $L_- = (L_+)^\dagger$, $L_z = Q_2$ and for the psedo-spin SU(2) by $L_+ = \eta'$ (see eq. (4)), $L_- = (\eta')^\dagger$, $L_z = 0.5(Q - N)$.

We define two quality factors, based on total spin and total pseudo-spin. The quality factor is defined as the time it takes for the total spin to decrease by a threshold factor (to be specified case-by-case) divided by the period of the unperturbed revivals.

The quality factors based on the total spin and pseudo-spin are shown in the bottom row in Fig. 1 for the same combinations of the initial state and the perturbation as those used for the fidelity-based quality factor calculations in the top row of Fig. 1. The controlled behaviour of the total (pseudo) spin-based quality factor allows to use it as an additional measure of MBS stability in presence of perturbations. This measure can be defined for any scars furnishing a unique representation of SU(2) (which may actually apply to some of the other known equally-spaced scars [27, 57, 65]).

While the quality factor based on the total spin has a transparent physical interpretation it can probably not be measured directly in experiment. One may however be able to extract information about its behaviour from the measurements of the projection of the total spin on at least two axes (Appendix Fig. 6 illustrates this).

Another possibility to detect MBS experimentally arises due to the characteristic correlations of the eta and zeta scar states given by the operators $G_O(i,j)$ and $G_U(i,j)$ defined in eqs. (6) and (5). The values of these correlators are known exactly analytically in the scar states and do not depend on the distance $(i-j)$ between the points where they are measured. For example for the zeta states we have finite expectation value [29]

$$G_U^{n\tilde{\zeta}} = \frac{1}{4} - \frac{n(N-n)}{2N(N-1)}, \qquad (27)$$

while when evaluated in any generic states outside of the zeta subspace in an unperturbed system we get near-zero values (see for instance Fig. 4 in the Supplementary Material of Ref. [29]). Similarly, in an unperturbed system (when MBS is exact) the $G_O$ operator is only substantially non-zero within the eta subspace (see Fig. 8 in Ref. [1]). Therefore the finite, non-zero values of these correlators can be used as a sensitive indicator of the state being inside or outside the respective scar subspace.

We will study the time evolution (unperturbed case shown in Appendix Fig. 6) of the expectation value of these correlators numerically and define the quality factors based on the (decay of the) amplitude of their oscillations. In experiment one could choose the nearest-neighbour $i$ and $j$ such that the measurement is local. Numerically we evaluate the value averaged over all possible choices of sites $i$ and $j$: $\frac{\sum_{i<j} G(i,j)}{N(N-1)/2}$.

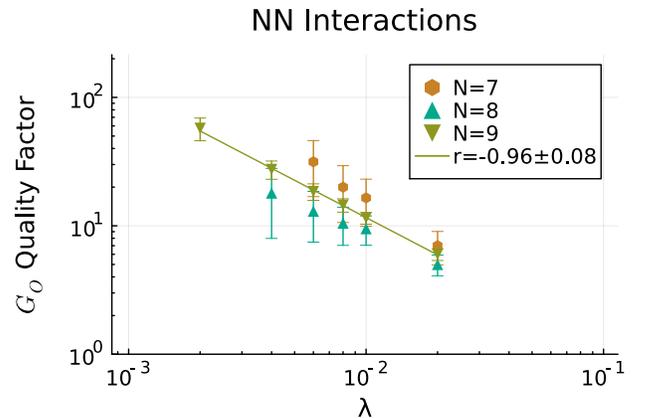

FIG. 2: Quality factor based on the superconducting $G_O$ correlations (6). The medians and their bootstrapped standard deviations are shown. The initial state is a superposition of the eta states, the perturbation is the nearest-neighbour density-density interaction (10 realizations). The threshold is set to 0.75. The fitted exponent $r$ for $N=9$ is indicated.

In Fig. 2 we show the quality factor based on the decay of the superconducting $G_O$ correlations (6) when starting from a mix of the eta states with the density-density perturbation. The power law dependence (26) we obtained for the fidelity-based quality factor continues to hold here as well.

Comparing the considered types of the quality factors for a fixed pair (initial state - perturbation type) over all the available data we conclude that the information they capture is qualitatively similar. An important difference however is that the fidelity-based quality factor measures the departure from the exact initial wavefunction while the quality factors based on physical quantities characterize the extent to which the time-evolved state preserves the observable properties of the whole scar subspace. For this reason one could expect the latter quality factors to be more stable than the fidelity-based one as a function of the system size. Indeed, in Fig. 2 for example we observe that the $G_O$-based quality factor does not strongly depend on the system size $N$.

For very small perturbations $\lambda \approx 10^{-5}$ the various quality factors are quantitatively very similar with the exception of the total spin-based quality factor typically yielding higher values. At higher perturbations ($\lambda \geq 10^{-3}$) we observe a tendency for the physical observables to be more stable compared to the fidelity. To see this one can compare the top and bottom rows in Fig. 1 where the top and bottom panels in the same column are using the same conditions and threshold. Further example of this is found for the case of random potential and eta states (data can be found in Appendix D) where we observe for $N = 9$ that the total pseudospin quality factor is about factor of 6 higher than the fidelity-based quality factor! Therefore the quality factors based on total (pseudo-) spin or the $G_O/G_U$ correlations are both preferable theoretically and are more accessible in potential experiments.

## VI. EXTRA SCAR STABILITY IN SMALL SYSTEMS

Small systems with $N$ between 4 and 50 are of particular interest for the exact numerical calculations and controlled simulations in near-term quantum simulators. As we show below, for small perturbations the time evolution that starts in the many-body scar subspace is qualitatively different from large $N$ counterparts and shows signs of additional stability.

This result is largely independent on the particular type of perturbation. To illustrate this, in this section, we replace the simple perturbations used so far (and listed in Tab. I) with a Hermitian random matrix from a GUE ensemble (referred to as a GUE perturbation) that couples a typical initial state to every eigenstate in the Hilbert space. Our exact unbiased numerical calculations will still be limited by $N = 9$.

The unusual behaviour is due to the fact that at small perturbations the original eigenstates are only slightly distorted. In particular, $N + 1$ of them mainly still have the character of the eta states and $N + 1$ further the character of the zeta states. This means that these "perturbed scars" are predominantly located in the original scar subspace (see Fig. 3a illustrating this). Thus an initial state $|\phi_0\rangle$ in the scar subspace only has a significant overlap with $N + 1$ eigenstates of the perturbed Hamiltonian and its time evolution can be approximately understood as the interference of these $N + 1$ eigenstates. Although in presence of perturbation these eigenenergies are no longer equally spaced for small $N$ this interference does not lead to complete averaging out. Even after an arbitrarily long time, the evolved state is mainly composed of the same $N + 1$ "perturbed scars", and the exact scar space is never left at small perturbations.

To make this statement more precise we derive in the Appendix E the expectation value of the projector $P$ to the scar subspace in the long-time limit, when all the phases of the interfering states can be assumed random. Starting from the

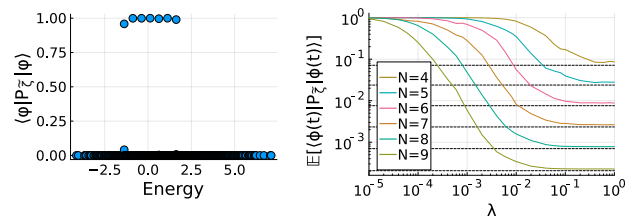

FIG. 3: Expectation value of the projector on the scar subspace. Left Panel: The expectation value is evaluated for every eigenstate of the perturbed Hamiltonian where a GUE perturbation of strength $10^{-4}$ was added to the unperturbed Hamiltonian (7). The projector on the zeta scar subspace $P_{\tilde{\zeta}} = \sum_n |n^{\tilde{\zeta}}\rangle \langle n^{\tilde{\zeta}}|$ was used. $N = 6$, open boundary conditions. The plot of the overlap of every eigenstate with a uniform superposition of the zeta states (not shown) looks identical. Right Panel: $t \to \infty$. The large-time expectation value of the projector to the zeta scar subspace for varying system sizes and GUE perturbation strengths calculated using eq. (28). The initial state is an equal-weight superposition of the zeta scars and we project to the zeta scar subspace. Dashed lines indicate the large-perturbation expectation values discussed in the text.

initial state $|\phi_0\rangle$ this expectation value is given by

$$\mathbb{E}_{t \to \infty} \left[ \langle \phi(t) | P | \phi(t) \rangle \right] = \sum_k \langle \varphi_k | P | \varphi_k \rangle |\langle \phi_0 | \varphi_k \rangle|^2, \quad (28)$$

where $|\varphi_k\rangle$ is an eigenstate of the perturbed Hamiltonian.

As we have seen (Fig. 3a) for small systems and small perturbations there are $N + 1$ eigenstates $|\varphi_j\rangle$ with $\langle \varphi_j | P_{\eta'} | \varphi_j \rangle \approx 1$ and $\sum_j |\langle \phi_0 | \varphi_j \rangle|^2 \approx 1$, while the other eigenstates have small projectors and overlaps. Inserting these approximations into (28) one can see that $\mathbb{E}_{t \to \infty} \left[ \langle \phi(t) | P_{\eta'} | \phi(t) \rangle \right] \approx 1$. The time-evolved state will continue to mainly reside within the exact scar subspace despite the fact that the GUE perturbation couples it to the entire Hilbert space. Furthermore, because the exact scar subspace is only $(N+1)$-dimensional the time-evolved state will approximately revisit the initial state $|\phi_0\rangle$ from time to time in small systems. An example of this time evolution at large times is given in the Appendix Fig. 7d.

In Fig. 3b we plot the value of the projector expectation value (28) for various system sizes and perturbation strengths which should provide a guidance to any experimental quantum simulations in small systems (the actual experimental perturbations will always be weaker than GUE). The large-perturbation value ($\lambda \leq 1$, see (13)), where no eigenstate subset has an increased overlap with the initial state, can be estimated as $\frac{N+1}{D_H}$, where $D_H$ is the Hilbert space dimension ($D_H = \binom{2N}{N}$ for the half-filling sector where the zeta states reside).

We observe that even with the increasing system sizes a sizeable projector expectation value is obtained for small enough perturbations.

## VII. MANY-BODY LOCALISATION

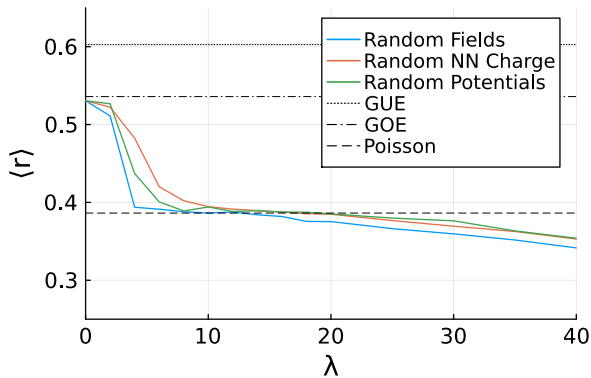

FIG. 4: The level statistic $\langle r \rangle$ computed numerically at half filling for different perturbations of varying strengths $\lambda$. Data shown is for $N = 9$ and open boundary conditions. A plateau at Poisson statistics is visible for all three perturbations, and it becomes more distinct with increasing system size (see Appendix Fig. 8).

A distinct regime of ergodicity breaking consistent with many-body localisation (note that some aspects of this phenomena in spin chains are a subject of ongoing debate [75, 76]) arises when the strength of the perturbations is increased above the energy scales given by the unperturbed system.

To identify the potential MBL region we tune the perturbation strength $\lambda$ and track the average (over the full spectrum) ratio $r$ level statistic that is known to be a sensitive measure for detecting the many-body localized phase [66]. $r_i$ is defined as the ratio of successive energy gaps $s_i = E_i - E_{i-1}$: $r_i = \frac{\min(s_i, s_{i+1})}{\max(s_i, s_{i+1})}$. It should be calculated upon resolving all the symmetries present in the system.

Expectation values for $r$ are known analytically [77]: $\langle r \rangle \approx 0.5359$ for the generalized orthogonal ensemble (GOE, real random matrices) and $\langle r \rangle \approx 0.6027$ for the generalized unitary ensemble (GUE, complex random matrices). These values signal that the system is ergodic. The Poisson value of $\approx 0.38629$ on the other hand is usually considered as a strong indicator of many-body localization and presence of the emerging associated conserved quantities.

The level statistics obtained numerically is shown in Fig. 4. With all the three perturbations considered in this work we find that for strengths between 8 and 20 the level statistics becomes approximately Poisson. In Appendix F we show that the distribution of $r$ indeed follows the one we expect from Poisson distributed energy gaps within this range. This observation suggests that the system is many-body localized. We should also keep in mind that the systems accessible to exact diagonalization are rather small and do exhibit finite-size effects. One of such effects is the deviation of the average level statistics parameter $\langle r \rangle$ away from the Poisson value observed at highest $\lambda$ values in Fig. 4. In the Appendix F by carefully analyzing the system size dependence we show that the relevant observables (including $\langle r \rangle$) do exhibit a strong trend towards typical MBL values with increasing system size also at higher values of $\lambda$.

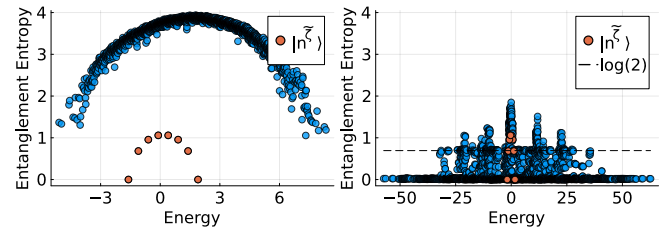

FIG. 5: Entanglement entropy computed for every half-filling eigenstate plotted as a function of that eigenstate's energy. We use $N = 7$ and open boundary conditions. Red data points indicate the zeta scars. Left Panel: unperturbed Hamiltonian (7). Right Panel: Random potential perturbation of strength $\lambda = 12$ is added.

MBL is typically expected to create states that are localized and therefore do not have significant entanglement. Numerically we study the bipartite entanglement entropy that arises if we cut our one-dimensional system in the middle. In the unperturbed regime the dependence of this entropy on energy forms a typical thermal arc (Fig. 5a) with the maximum corresponding to infinite temperature and reproduces the results previously known in this system [1]. The situation is drastically different at high perturbations as shown in Fig. 5b in presence of a random potential (results are similar for other perturbation types). Most of the eigenstates here have vanishing entanglement which is consistent with them being fully localised. The entanglement of the remaining states while finite is significantly lower than in the unperturbed system.

Careful analysis of the numerical data including the wave-functions suggests that the finite entanglement arises due to the fact that the randomized perturbations we use here (chosen as the most relevant impurities in a potential many-body scar experiment) do not couple to all the local degrees of freedom. The random on-site potentials (Fig. 5b) for example would fully localize the density on each site at $\lambda = \infty$. They are however insensitive to the spin configuration and for a Hamiltonian made of the perturbation only the states with identical density but different spin configurations would make a degenerate subspace. The number of states in such a subspace is 2 to the power given by the number of half-filled sites in a density configuration. The density configurations with the most half-filled sites lead to the highest finite entanglement in the perturbed system seen as almost vertical clusters of states in Fig. 5b. At high perturbation strengths the interacting part $H_h^{tJU}$ (7) of the Hamiltonian can be thought to either split the degenerate levels of the perturbation term or couple its levels very close in energy. This leads to states with finite entanglement and in particular to $\log(2)$ entanglement for the cases when the resulting state is a superposition of just two degenerate eigenstates of the perturbation term.

Even though the number of finite-entanglement states grows with sistem size as explained above this growth is outpaced by the $4^N$ growth of the full Hilbert space and as we show in the Appendix F the fraction of the delocalized states vanishes as the system size is increased.

Further evidence for the non-thermal behaviour in the strong-perturbation regime is provided by the time evolution of the entanglement entropy and the out-of-time-order correlators (OTOC) [78] also discussed in Appendix F.

Overall our finite-size numerical results clearly indicate that at strong perturbation strengths the system is non-thermal and the number of states deviating from the thermal behaviour is extensive as opposed to the many-body scar regime. While such numerics can not in principle strictly prove the existence of the MBL phase our observations are consistent with many-body localization once the finite-size effects are taken into consideration.

## VIII. OUTLOOK AND DISCUSSION

The zeta scar states are completely insensitive to the on-site chemical potential while the eta-pairing scars are insensitive to the on-site magnetic field. These two scar families remain stable under magnetic fields and chemical potentials respectively in the first order of the perturbation theory. Both these results are due to the high symmetry of the two scar families and are Hamiltonian-independent as long as the Hamiltonian is a part of a large class of models [1] that support eta and zeta states as exact scars.

The quality factor based on the decay of the total quadratic Casimir operator of the underlying SU(2) symmetry group is a natural measure of scar stability that should be applicable to any equispaced scars. In the particular case of scars in spin-1/2 fermionic systems we further propose that the stability of scars can be assessed based on the experimentally measurable dynamics of their characteristic ODLRO correlators.

These results should provide guidance for the future experimental observations of many-body scars in fermionic systems and facilitate the development of their quantum information processing applications.

Another motivation for the future experimental studies is the presence of two very different ergodicity-breaking phenomena in the same system: while few many-body scars break ergodicity at zero or small perturbation an extensive fraction of the Hilbert space exhibits non-thermal behaviour at strong disorder. We show that one can go from one regime to the other by simply tuning the strength of a local impurity perturbation.

It is conceivable that one can think about photo-induced superconductivity [79–81] in terms of effectively preparing a state that has a finite projection on the scar subspace of eta-pairing states. It would be interesting to study the relation between the $G_O$-based quality factors and the decay times we discussed here and the time scales observed in experiments [79, 80, 82].

## IX. ACKNOWLEDGEMENTS

We thank Alexey Milekhin, Fedor Popov, Dmitry Abanin and Manfred Sigrist for many useful discussions. Some of the results presented here are from Patrice Kolb's ETH Zurich MSc Thesis (Sep 2022). We acknowledge access to Piz Daint at the Swiss National Supercomputing Centre, Switzerland under the ETHZ's share with the project ID ETH8. We also thank Marina Marinkovic for assisting us with obtaining the said access.

## Appendix A: Perturbation theory

### 1. Random potentials on eta scars

Let us derive the expression

$$\Delta E_n^{\eta'}(\lambda) = \langle n^\eta |' \lambda n_l | n^\eta \rangle' = \frac{2n}{N}\lambda. \tag{A1}$$

Let us rewrite the $|n^\eta\rangle'$ states (4):

$$\begin{aligned}
|n^\eta\rangle' &= \mathcal{N}_{N,n} \left( \sum_j (-1)^j c_{j\uparrow}^\dagger c_{j\downarrow}^\dagger \right)^n |0\rangle \\
&= \mathcal{N}_{N,n} \sum_{j_1,\ldots,j_n} \prod_{k=1}^n (-1)^{j_k} c_{j_k\uparrow}^\dagger c_{j_k\downarrow}^\dagger |0\rangle \\
&= \mathcal{N}_{N,n} \sum_{\substack{j_1,\ldots,j_n \\ \text{all different}}} |j_1,\ldots,j_n\rangle \\
\text{with} \quad &|j_1,\ldots,j_n\rangle := \prod_{k=1}^n (-1)^{j_k} c_{j_k\uparrow}^\dagger c_{j_k\downarrow}^\dagger |0\rangle
\end{aligned} \tag{A2}$$

Note that the order of the excitations does not matter since $[c_{j\uparrow}^\dagger c_{j\downarrow}^\dagger, c_{k\uparrow}^\dagger c_{k\downarrow}^\dagger] = 0 \quad \forall j,k$. We can restrict the sum to all different $j_1,\ldots,j_n$ because $(c_{j\uparrow}^\dagger c_{j\downarrow}^\dagger)^2 = 0$.

We can now apply the on-site potential operator $\lambda n_l$ to the scar. It provides a factor $2\lambda$ to states where $l$ is occupied and annihilates the other ones:

$$\lambda n_l |n^\eta\rangle' = 2\lambda n \, \mathcal{N}_{N,n} \sum_{\substack{j_1,\ldots,j_{n-1} \neq l \\ \text{all different}}} |j_1,\ldots,j_{n-1},l\rangle \tag{A3}$$

The factor $n$ comes from the $n$ possible $j_k$ which could correspond to $l$. Finally, we can apply $\langle n^\eta|'$ on the left to obtain the first order energy correction:

$$\langle n^\eta |' \lambda n_l | n^\eta \rangle' =$$
$$2\lambda n \, \mathcal{N}_{N,n}^2 \left( \sum_{\substack{j_1,\ldots,j_n \\ \text{all different}}} \langle j_1,\ldots,j_n| \right) \left( \sum_{\substack{j_1,\ldots,j_{n-1} \\ \text{all different}}} |j_1,\ldots,j_{n-1},l\rangle \right) \tag{A4}$$

Note that

$$\langle j'_1, ..., j'_n | j_1, ..., j_n \rangle = \begin{cases} 1 \text{ if } j'_1, ..., j'_n \text{ are a permutation of } j_1, ..., j_n \\ 0 \text{ otherwise.} \end{cases} \quad (A5)$$

For evaluating (A4) we can thus simply count the number of contributing combinations of states between the two brackets. The right bracket in equation (A4) contains $\frac{(N-1)!}{((N-1)-(n-1))!} = \frac{(N-1)!}{(N-n)!}$ terms, corresponding to the number of ways to choose $n-1$ different values in $\{1, ..., N\} \setminus \{l\}$ to assign to $j_1, ..., j_{n-1}$. Each term in the right bracket is matched by $n!$ terms on the left hand side: The values $j_1, ..., j_n$ must match the indices in the right bracket to arrive at the same occupation number state, they can be ordered in $n!$ different ways. The bracket product thus evaluates to $n! \cdot \frac{(N-1)!}{(N-n)!}$. Inserting the expression for the normalization constant $\mathcal{N}_{N,n}$ we arrive at the first order correction

$$\langle n^\eta |' \lambda n_l | n^\eta \rangle' = 2\lambda n \, \frac{(N-n)!}{N!n!} n! \frac{(N-1)!}{(N-n)!} = \frac{2n}{N}\lambda. \quad (A6)$$

With random potentials on each site, $V = \sum_{l=1}^{N} \lambda_l n_l$, we get

$$\langle n^\eta |' \sum_{l=1}^{N} \lambda_l n_l | n^\eta \rangle' = \sum_{l=1}^{N} \langle n^\eta |' \lambda_l n_l | n^\eta \rangle' = \frac{2n}{N} \sum_{l=1}^{N} \lambda_l \quad (A7)$$

which still keeps the scar energies equispaced.

### 2. Density-density interactions on eta scars

Given a charge interaction between two sites $l$ and $m$, the first order energy correction on the $|n^\eta\rangle'$ scars is given by $\langle n^\eta |' \lambda n_l n_m | n^\eta \rangle'$. It is again helpful to express the scars as $|n^\eta\rangle' = \mathcal{N}_{N,n} \sum_{\substack{j_1, ..., j_n \\ \text{all different}}} |j_1, ..., j_n\rangle$ as in section A 1. Applying the operator $\lambda n_l n_m$ annihilates all states where either $l$ or $m$ (or both) are not occupied. The remaining states receive a factor of $4\lambda$:

$$\lambda n_l n_m |n^\eta\rangle' = 4\lambda \mathcal{N}_{N,n} n(n-1) \sum_{\substack{j_1, ..., j_{n-2} \\ \text{all different and } \neq l,m}} |j_1, ..., j_{n-2}, l, m\rangle \quad (A8)$$

The factor $n(n-1)$ corresponds to the number of possible pairs of indices in $\{1, ..., n\}$ one could assign to $l$ and $m$. The sum contains $\frac{(N-2)!}{((N-2)-(n-2))!} = \frac{(N-2)!}{(N-n)!}$ terms, corresponding to the number of possibilities to assign the remaining $N-2$ sites to $j_1, ..., j_{n-2}$ without repetition. Applying $\langle n^\eta |'$ on the left hand side, each term of the sum in (A8) is matched by the terms in $\langle n^\eta |'$ where the same sites are occupied. There are $n!$ such terms due to the different possible orderings. Making use of the orthogonality of the states $|j_1, ..., j_n\rangle$ (A5) we can thus determine the first order correction as

$$\langle n^\eta |' \lambda n_l n_m | n^\eta \rangle' = 4\lambda \mathcal{N}_{N,n}^2 \cdot n(n-1) \cdot \frac{(N-2)!}{(N-n)!} \cdot n!$$
$$= 4\lambda \frac{(N-n)!}{N!n!} \cdot n(n-1) \cdot \frac{(N-2)!}{(N-n)!} \cdot n!$$
$$= 4\lambda \frac{n(n-1)}{N(N-1)}. \quad (A9)$$

Note that this first order correction is quadratic in $n$. This breaks the equal spacings: The energies shifted by the first order are given by (see (11), (A9))

$$E_n^{\eta'}(\lambda = 0) + \Delta E_n^{\eta'}(\lambda) = (U - 2\mu)n + 4\lambda \frac{n(n-1)}{N(N-1)} \quad (A10)$$

leading to $n$-dependent energy spacings:

$$(E_{n+1}^{\eta'}(\lambda=0) + \Delta E_{n+1}^{\eta'}(\lambda)) - (E_n^{\eta'}(\lambda=0) + \Delta E_n^{\eta'}(\lambda))$$
$$= U - 2\mu + 4\lambda \frac{2n}{N(N-1)}. \quad (A11)$$

The $|n^\eta\rangle'$ scar states are thus unstable under density-density interactions already in first order.

In a system with significant density-density interactions one could in principle assure that the energies corrected in first order still support revivals by tuning the parameters such that $U - 2\mu = k \cdot \Delta\epsilon$ with $k \in \mathbb{Z}$ and $\Delta\epsilon = \frac{4\lambda}{N(N-1)}$. Inserting this into (A10) we then find that $E_n^{\eta'}(\lambda=0) + \Delta E_n^{\eta'}(\lambda) = (kn + n(n-1)) \cdot \Delta\epsilon$. In a tuned system, the corrected energies are thus integer multiples of a common $\Delta\epsilon$ and support revivals with a period of $T = \frac{2\pi}{\Delta\epsilon}$.

As in the previous sections we can directly generalize these results to arbitrary density-density interactions between all sites.

## Appendix B: Quality factor dependence on the perturbation strength

We consider an arbitrary scarred initial state $|\phi_0\rangle = \sum_{n=0}^{N} a_n |n\rangle$, where $|n\rangle$ are scar states (from either eta or zeta families) with equispaced energies $\epsilon_n = \epsilon_0 + n \cdot \Delta\epsilon$. In the unperturbed system the fidelity returns to 1 at all *peak times* $t_l = lh/\Delta\epsilon, l \in \mathbb{N}$:

$$|\langle \phi_0 | \phi(t_l) \rangle|^2 = 1. \quad (B1)$$

Let us investigate how the fidelity peaks change if we perturb the Hamiltonian to $H_\lambda = H + \lambda V$. The time evolution is then given by c

$$|\phi(t)\rangle = e^{-\frac{i}{\hbar}H_\lambda t}|\phi_0\rangle = \sum_{n=0}^{N} a_n e^{-\frac{i}{\hbar}H_\lambda t}|n\rangle. \quad (B2)$$

We can write

$$e^{-\frac{i}{\hbar}H_\lambda t}|n\rangle = \sum_k c_{n,k}(t) e^{-\frac{i}{\hbar}\epsilon_k t}|k\rangle, \quad \text{(B3)}$$

where $k$ moves over all eigenstates of $H$. This leads to the fidelities

$$|\langle \phi_0|\phi(t)\rangle|^2 = \left|\sum_{n,m=0}^N a_m^* a_n c_{n,m}(t) e^{-\frac{i}{\hbar}\epsilon_m t}\right|^2. \quad \text{(B4)}$$

At $\lambda = 0$ we have $c_{n,m}(t) = \delta_{nm}$, which using the fact that at the peak times $e^{-\frac{i}{\hbar}\epsilon_n t_l} = e^{-2\pi i \frac{\epsilon_0}{\Delta\epsilon}l}$ (independent of $n$) yields $|\langle \phi_0|\phi(t_l)\rangle|^2 = 1$.

At small $\lambda > 0$ we can expand $c_{n,m}(t)$ using time dependent perturbation theory. Up to second order we have (eq. (5.369) in [83])

$$c_{n,n}(t) = \exp\left(-\frac{i}{\hbar}(\lambda \Delta_n^{(1)} + \lambda^2 \Delta_n^{(2)})t\right) \quad \text{(B5)}$$

with $\lambda \Delta_n^{(1)} = \langle n|\lambda V|n\rangle$ the first order energy correction to $\epsilon_n$, and (eq. (47) in [84])

$$c_{n,m\neq n}(t_l) = \left(\frac{\lambda}{i\hbar}\right)^2 \left(\frac{V_{mn}V_{nn}}{i\omega_{mn}}t_l - \frac{V_{mm}V_{mn}}{i\omega_{mn}}t_l - \sum_{k\neq n,m}\frac{V_{mk}V_{kn}}{\omega_{mk}\omega_{kn}}(1 - e^{i\omega_{mk}t_l})\right) \quad \text{(B6)}$$

with $V_{ab} := \langle a|V|b\rangle$, $\omega_{ab} := (\epsilon_a - \epsilon_b)/\hbar$. $k$ again labels all eigenstates of $H$. Here we used the fact that at the peak times $e^{i\omega_{nm}t_l} = 1$.

In first order of $\lambda$ we can neglect $c_{n,m\neq n}(t_l)$ and approximate

$$c_{n,n}(t) \approx \exp\left(-\frac{i}{\hbar}\lambda \Delta_n^{(1)} t\right), \quad \text{(B7)}$$

leading to the fidelity

$$|\langle \phi_0|\phi(t)\rangle|^2 \approx \left|\sum_{n=0}^N |a_n|^2 \exp\left(-\frac{i}{\hbar}(\epsilon_n + \lambda \Delta_n^{(1)})t\right)\right|^2. \quad \text{(B8)}$$

At the peak times ($e^{-\frac{i}{\hbar}\epsilon_n t_l} = e^{-2\pi i \frac{\epsilon_0}{\Delta\epsilon}l}$) it evaluates to

$$|\langle \phi_0|\phi(t_l)\rangle|^2 \approx \left|\sum_{n=0}^N |a_n|^2 \exp\left(-\frac{i}{\hbar}\lambda \Delta_n^{(1)} t\right)\right|^2. \quad \text{(B9)}$$

The quality factor is proportional to the time where this quantity reaches a certain threshold. Since the time is scaled by $\lambda$ the quality factor will generally follow a power law with respect to $\lambda$ in first order:

$$Q \propto \lambda^{-1} \quad \text{(B10)}$$

However, notice that (B8) corresponds to the unperturbed expression up to the fact that the scar's energies have been corrected in first order. Thus, if the first order correction leaves the energies equispaced with gaps $\Delta\epsilon + \lambda c$, $c \in \mathbb{R}$, the fidelity will still return to 1 in first order, albeit at peak times $t'_l = h/(\Delta\epsilon + \lambda c)$. In these cases let us consider the second order in $\lambda$:

$$|\langle\phi_0|\phi(t'_l)\rangle|^2 \approx$$
$$\left|\sum_{n=0}^N |a_n|^2 e^{-\frac{i}{\hbar}\lambda^2 \Delta_n^{(2)} t'_l} + \sum_{\substack{n,m=0 \\ n\neq m}}^N a_m^* a_n c_{n,m}(t'_l)\right|^2 \quad \text{(B11)}$$

In the first term here and in the first two terms entering $c_{n,m}$ in (B6) the time is multiplied by $\lambda^2$ and we can anticipate (keeping in mind that the magnitude of the third term in (B6) is bound w.r.t. time $t$) this to result in the $Q \propto \lambda^{-2}$ dependence. The expression for $c_{n,m}(t_l)$ (B6) has been calculated with respect to the unperturbed peak times $t_l$. Since the peak time correction is of order $\lambda$ we can use the same expression at $t'_l$ in second order. There is a subtlety with this argument: The smaller $\lambda$, the longer it takes until the threshold defining the quality factor is crossed and the larger times we need to consider. We need to assume that for the range of $\lambda$ we consider the threshold is chosen high enough.

Particularly the last term of (B6) deserves some attention as it is not linear in $t'_l$. Note however that due to the denominator $\omega_{mk}\omega_{kn}$ the contributions of eigenstates $|k\rangle$ with energies close to either $|n\rangle$ or $|m\rangle$ dominate the sum. If $\epsilon_k \approx \epsilon_n$ we have $\hbar\omega_{mk} \approx (m-n)\Delta\epsilon$ and if $\epsilon_k \approx \epsilon_m$, $\omega_{mk} \approx 0$. Motivated by this, let us thus assume a threshold high enough such that $e^{i\omega_{mk}t'_l} \approx 1$ in both cases and $(1 - e^{i\omega_{mk}t'_l})$ can be approximated as linear in $t'_l$. In this approximation we have $c_{n,m\neq n}(t'_l) \propto \lambda^2 t'_l$, and the time is scaled by $\lambda^2$ in all factors of (B11), leading to

$$Q \propto \lambda^{-2}. \quad \text{(B12)}$$

In summary, the quality factor is related to the perturbation strength $\lambda$ through a power law at small $\lambda$ and large enough thresholds. The first order corrections we calculated in section IV A control whether the exponent is $-1$ or $-2$.

**Appendix C: Dynamics in the unperturbed system**

If we start from a half-filled state where every spin is up (which is a linear combination of the zeta scar states) we observe that the total spin projections on the $Z$ and $X$ axis oscillate as shown in Fig. 6. This can be understood using a simple model of a magnetic moment $m$ in a uniform magnetic field. The magnetic moment carried by the initial state is along the $Z$ axis. The $Q_2$ term in the unperturbed Hamiltonian (7) is a magnetic field along the negative $Y$ axis. The torque exerted by this external magnetic field is given by $\tau = m \times B_{ext}$. It determines the rate of change of the magnetic moment and leads to its precession in the $X - Z$ plane as is observed in Fig. 6.

Note that this observation may be quite general: the majority of the known many-body scar states actually forms an equally-spaced in energy tower. One could potentially interpret it as a unique representation of an SU(2) group split by an effective magnetic field. The dynamics of a state from the scar subspace then corresponds to a precession analogous to the one we observe here.

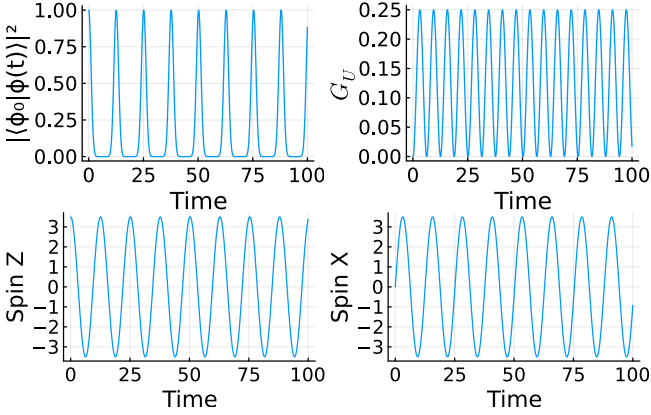

FIG. 6: Time evolution of observables in an unperturbed system with open boundary conditions and $N = 7$. The initial state is a product "all up" half-filling state which is a superposition of the zeta scars. The observables plotted are the fidelity to the initial wavefunction (top left), the expectation value of the $G_U$ correlator (5) (top right), the total spin projection on the $Z$ axis (bottom left) and the total spin projection on the $X$ axis (bottom right).

As we see in the plot of the total spin projection on the $Z$ axis the initial state oscillates between the all up state and the all down state. Both of these states have a vanishing $G_U$ expectation value which leads to a doubled frequency in the $G_U$ oscillations.

### Appendix D: Numerical data for eta states and on-site random chemical potential

Fig. 7a)-c) shows the quality factor data obtained with the random on-site potentials acting as a perturbation. As illustrated in the top left panel (quality factor based on the wave-function fidelity) the behaviour predicted by the perturbation theory is only observed for $\lambda \leq 10^{-2}$.

With this exception the data obtained for the on-site potential shows the same qualitative behaviour as the two other perturbation types and thus supports the conclusions made in the main text.

### Appendix E: Infinite time behavior

In order to quantify how closely the evolving states remain to the exact scar space, let us calculate the expectation value of the scar projector expectation value after a long time where the phases are completely randomized. That is, let us find $\mathbb{E}\left[\langle\phi(t)|P|\phi(t)\rangle\right]$ for large $t$ with $P = \sum_n |n\rangle\langle n|$ the projector to the space spanned by exact scar states $|n\rangle$. Inserting the

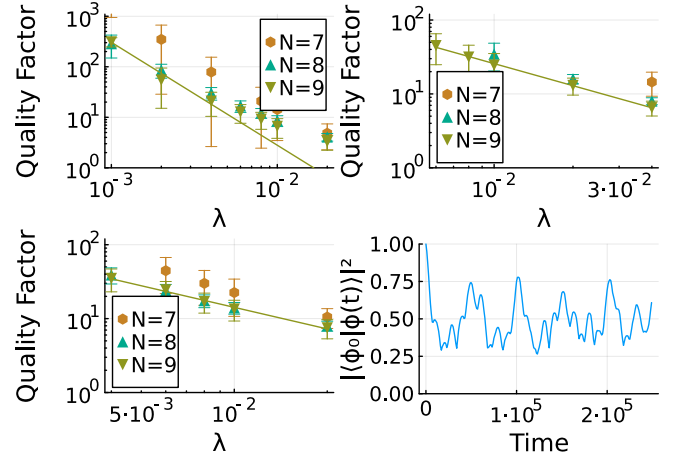

FIG. 7: Top and Bottom Left Panels: The quality factors with the perturbation of varying strengths $\lambda$ given by the random (10 realizations thereof) on-site potential. The medians and their bootstrapped standard deviations are shown. The fitted exponent $r$ is provided for $N = 9$. The initial state is an even superposition of all eta scars. Top Left: Overlap-based quality factor, the threshold is 0.9. Only data with $\lambda \leq 4 \cdot 10^{-3}$ was fitted. $r = -2.04$. Top Right Panel: Quality factor based on the total pseudospin. $r = -0.99$ (fit beyond the range of applicability of the perturbation theory). The threshold is 0.75. Bottom Left Panel: Quality factor based on the averaged (over sites where the correlation is measured) $G_O$ correlator. $r = -0.97$ (fit beyond the range of applicability of the perturbation theory). The threshold is 0.75. Bottom Right Panel: Long-time time evolution of the revivals amplitude in a small system with $N = 7$ sites where the GUE perturbation (Hermitian random matrix) with strength $\lambda = 10^{-4}$ is used. The initial state is an "all up" product state which is a superposition of the zeta states. The time evolution is performed within the half-filling sector.

definition of $P$, we obtain

$$\mathbb{E}\left[\langle\phi(t)|P|\phi(t)\rangle\right] = \sum_{n=0}^{N} \mathbb{E}\left[\langle\phi(t)|n\rangle\langle n|\phi(t)\rangle\right]$$
$$= \sum_{n=0}^{N} \mathbb{E}\left[|\langle n|\phi(t)\rangle|^2\right]$$
$$= \sum_{n=0}^{N} \mathbb{E}\left[\text{Re}^2\langle n|\phi(t)\rangle\right] + \mathbb{E}\left[\text{Im}^2\langle n|\phi(t)\rangle\right]. \tag{E1}$$

Expressing both the scar states and the initial condition in terms of the eigenstates $|\varphi_k\rangle$ of the system,

$$|n\rangle = \sum_k a_k^n |\varphi_k\rangle, \quad |\phi_0\rangle = \sum_k c_k |\varphi_k\rangle \to |\phi(t)\rangle \tag{E2}$$
$$= \sum_k c_k e^{-iE_k t} |\varphi_k\rangle,$$

we can write

$$\langle n|\phi(t)\rangle = \sum_k \bar{a}_k^n c_k e^{-iE_k t}. \quad (E3)$$

After long times, $-E_k t$ can be treated as uniformly distributed and uncorrelated phases $\alpha_k \in [0, 2\pi)$. Absorbing the phases of $\bar{a}_k^n c_k$ into $\alpha_k$ we get

$$\langle n|\phi(t)\rangle = \sum_k |a_k^n||c_k|e^{i\alpha_k}, \quad (E4)$$

$$\mathrm{Re}\,\langle n|\phi(t)\rangle = \sum_k |a_k^n||c_k|\cos\alpha_k.$$

We can now calculate $\mathbb{E}\left[\mathrm{Re}^2\langle n|\phi(t)\rangle\right]$ using the identity

$$\mathrm{Var}[X] = \mathbb{E}[X^2] - (\mathbb{E}[X])^2 \quad (E5)$$
$$\to \mathbb{E}[X^2] = \mathrm{Var}[X] + (\mathbb{E}[X])^2$$

with $X = \mathrm{Re}\,\langle n|\phi(t)\rangle$. Looking at (E4) and remembering that $\alpha_k$ is uniformly distributed, it is clear that $\mathbb{E}\left[\mathrm{Re}\,\langle n|\phi(t)\rangle\right] = 0$ as the expectation value of each term vanishes individually. We can obtain the variance using $\mathrm{Var}[X+Y] = \mathrm{Var}[X] + \mathrm{Var}[Y]$ and $\mathrm{Var}[\lambda X] = \lambda^2 \mathrm{Var}[X]$ together with (E4),

$$\mathrm{Var}\left[\mathrm{Re}\,\langle n|\phi(t)\rangle\right] = \sum_k |a_k^n|^2|c_k|^2 \mathrm{Var}[\cos\alpha_k]. \quad (E6)$$

The variance of $\cos\alpha_k$ with an uniformly distributed $\alpha_k$ can be found by again using (E5):

$$\mathrm{Var}[\cos\alpha_k] = \mathbb{E}[\cos^2\alpha_k] - (\mathbb{E}[\cos\alpha_k])^2$$
$$\mathbb{E}[\cos^2\alpha_k] = \frac{1}{2\pi}\int_0^{2\pi} d\alpha_k \cos^2\alpha_k = \frac{1}{2}$$
$$\mathbb{E}[\cos\alpha_k] = \frac{1}{2\pi}\int_0^{2\pi} d\alpha_k \cos\alpha_k = 0$$
$$\Rightarrow \mathrm{Var}[\cos\alpha_k] = \frac{1}{2} \quad (E7)$$

The first integral can be obtained using partial integration. Combining (E5), (E6), and (E7) we obtain

$$\mathbb{E}\left[\mathrm{Re}^2\langle n|\phi(t)\rangle\right] = \frac{1}{2}\sum_k |a_k^n|^2|c_k|^2. \quad (E8)$$

For the imaginary part we get the same result since $\mathrm{Im}\,e^{i\alpha_k} = \sin\alpha_k$ only differs by a constant phase shift. We can insert this into (E1) to arrive at the projector expectation value:

$$\mathbb{E}\left[\langle\phi(t)|P|\phi(t)\rangle\right] = \sum_{n=0}^N \sum_k |a_k^n|^2|c_k|^2 \quad (E9)$$
$$= \sum_k \langle\varphi_k|P|\varphi_k\rangle|\langle\phi_0|\varphi_k\rangle|^2$$

**Appendix F: MBL**

We first consider the system size dependence of the level statistics parameter $r$. From the early literature [66] on it is known that in small systems deviations from the reference ensemble values are to be expected.

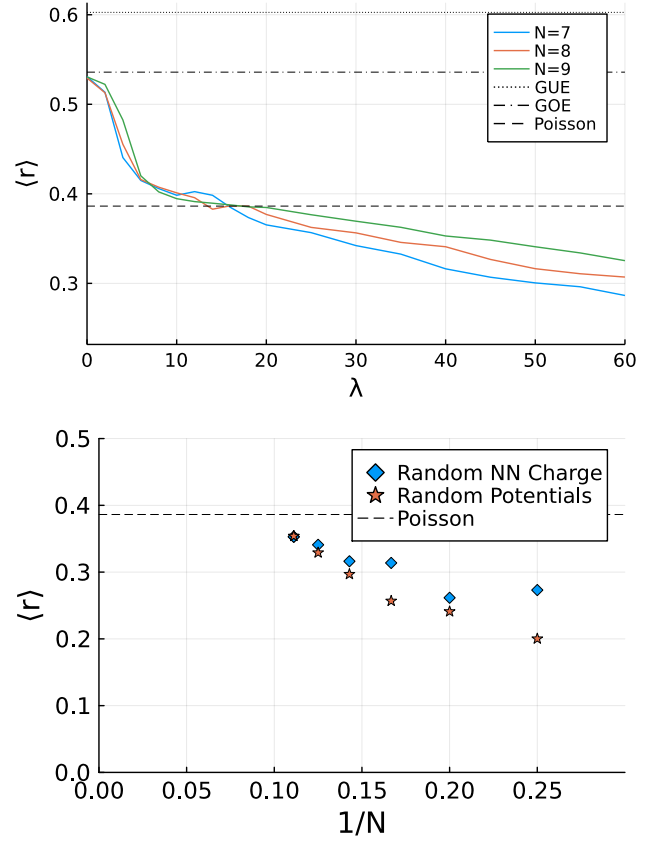

FIG. 8: Top Panel: Average level statistics $\langle r \rangle$ within the half-filling sector as a function of the strength of the nearest-neighbour density-density perturbation. Horizontal lines indicate the reference values for random matrix ensembles (GOE and GUE) and the Poisson statistics. Bottom Panel: System size dependence of the average level statistics parameter $\langle r \rangle$ for $\lambda = 40$ and two different types of perturbations.

Fig. 8a shows the perturbation strength dependence of the average level statistics parameter $r$ for three different system sizes. We observe that the plateau in the vicinity of the Poisson reference value becomes more pronounced with increasing system size. Poisson statistics is expected in the many-body localized regime as a result of the emergence of a large number of conservation laws due to disorder. Besides the average value of $r$ we also examined its distribution within the Poisson plateau shown in Fig. 9 and found excellent agreement with the distribution expected in the MBL regime [66]. We notice that at the higher perturbation strengths ($\lambda > 20$) the $r$ dependence moves closer to the Poisson value as the system size is increased thereby bringing the dependence closer to the typical curve one would expect for MBL. A plausible reason for the deviation from the Poisson value at extremely strong perturbations is as follows: At large $\lambda$ the Hamiltonian $H_\lambda = H + \lambda V$ can be seen as $\lambda V$ with a small perturbation $H$. $\lambda V$ typically has highly degenerate eigenenergies, both the random potentials and the density-density interactions do not distinguish between the possible spins of sites occupied by one particle for example. $H$ weakly splits the degenerate energy levels, but they will remain crowded around the eigenenergies of $\lambda V$. This means that we have an effective "level attraction" indicated by a value of $r$ smaller than what we expect from Poisson distributed energy spacings.

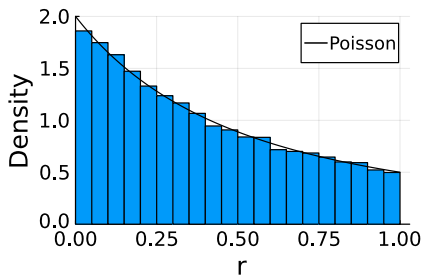

FIG. 9: Distribution of the $r$ values at half filling in a chain of $N = 9$ sites under a random potential perturbation of strength $\lambda = 20$. The distribution expected from Poisson distributed energy spacings ($2/(1 + r)^2$, see [66]) is overlayed.

In Fig. 8b we investigate the system size dependence of $\langle r \rangle$ in more detail and plot it for two different perturbation types as a function of $1/N$ for $\lambda = 40$ - the value where the deviation from the Poisson reference in Figs. 4 and 8a may appear significant. In both cases we observe a clear strong trend towards the Poisson value with growing system size. Therefore, regardless of the reason that causes the $r$ dependence to drift lower than the Poisson value we expect this effect to be negligible in large enough systems such that the MBL region we called "Poisson plateau" so far actually extends to arbitrarily high interaction strengths.

We now turn to the finite-size effects observed in the entanglement entropy data. One can see in Fig. 5 that although most of the states have zero or drastically reduced entanglement in presence of strong disorder, some states remain fairly delocalized with finite entanglement entropy. To quantify this effect we count the number of all the states with the entanglement entropy larger than 0.5 and calculate the fraction of such states in the whole Hilbert space for random potential disorders of strengths within the Poisson plateau of the $\langle r \rangle$ dependence. The results in Fig. 10a show a strong trend for the fraction of the delocalized states to decrease with the system size and we expect their contribution to be vanishing at large enough sizes.

The time evolution of the entanglement entropy starting from a random product state can be used as another diagnostic of MBL [67]. We exhibit in Fig. 11 the entanglement time evolution for $N = 7$ with and without (random on-site potential) disorder. In case of the unperturbed system we observe rapid growth of entanglement as expected in an ergodic, interacting and thermalizing system. The result is clearly non-thermal in presence of strong perturbations (within the MBL regime $\lambda = 12$ judging by the level statistics) - the entangle-ment growth is much slower although it is not logarithmic. It is possible that the truly logarithmic growth (expected for MBL [67]) could only be seen in larger system sizes as we do observe some other MBL signatures discussed above becoming more clear with increasing system size.

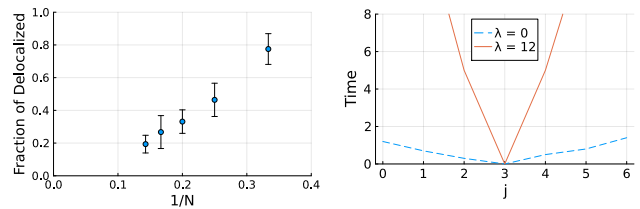

FIG. 10: Left Panel: Fraction of the delocalized (entanglement entropy larger than 0.5) half-filling states as a function of the inverse system size. We show the median and its bootstrapped standard deviation over 10 realizations of random on-site potentials and choose a perturbation strength within the Poisson plateau for each realization based on $\langle r \rangle$. Right Panel: Evolution of the OTOC (F1) starting from a random product state. We fix one of the site indexes $k = 4$. Perturbation used is the random potential.

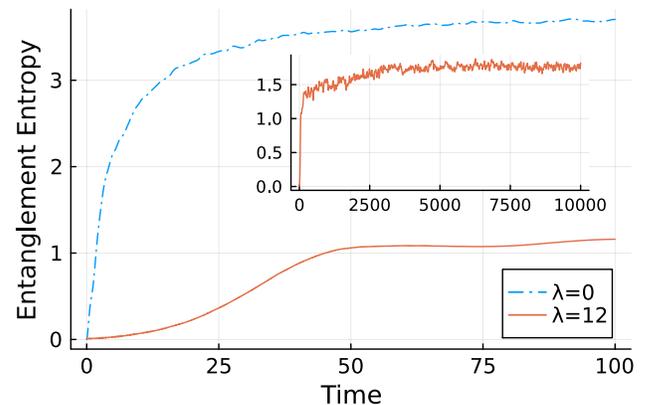

FIG. 11: Entanglement entropy time evolution starting from a random product state. The cut is made in the middle of the 1D chain. $\lambda = 0$ corresponds to the unperturbed Hamiltonian (7). $\lambda = 12$ data is for the same Hamiltonian where strong random chemical potential has been added corresponding to the middle of the MBL plateau seen in level statistics (Fig. 4). The inset shows the entanglement time evolution in the perturbed case on a much longer time scale.

The out-of-time-order correlator (OTOC) [78] is a concept that can be used to characterize the effective speed with which a perturbation caused by a local operator spreads in space. Given two operators $V_j$ and $W_k$ acting on sites $k$ and $j$ respectively and an initial state $|\psi\rangle$ it is given by (mind a slight notation change compared to Ref. [78])

$$F_{kj}(t) = \langle \psi | W_k e^{iHt} V_j e^{-iHt} W_k e^{iHt} V_j e^{-iHt} | \psi \rangle. \quad \text{(F1)}$$

We study the particular OTOC specified by $W_k = \sigma_k^2$ (Pauli Y matrix acting on the spin on site k) and $V_j = \sigma_j^3$ (Pauli Z

matrix acting on the spin on site j). The initial state $|\psi\rangle$ is a random product state. To study the spread of a perturbation we fix $k = 4$. As long as the information at $k = 4$ has not reached $j$, the OTOC $F_{kj}(t)$ remains close to 1. Once the operator $W_k$ affects $j$ $F_{kj}(t)$ is decreased. Tracking the time and position where such a decrease happens one can draw an approximate "light cone" of the information spreading.

The right panel of Fig. 10 shows this light cone for the perturbed (in the MBL regime) and unperturbed systems. With no perturbation, the information reaches the outermost site in 1.3 time units, leading to a velocity of 2.3 sites per time unit. With the perturbation the propagation takes 19 time units corresponding to the velocity of 0.16.

We observe that the information spreads an order of magnitude slower in the perturbed system which can be considered as another manifestation of it being localized.

Summarizing we can say that all the data we have undoubtedly suggests that in contrast to the many-body scars the number of states breaking ergodicity at high perturbation strength is extensive. Our data is also consistent with many-body localisation occurring in this regime although finite-size effects are still strong for the sizes accessible to exact diagonalization.